\documentclass[aps,prl,showpacs, amsmath,
  amssymb,reprint,groupedaddress]{revtex4-1}
\usepackage{color}
\usepackage{graphicx}
\begin{document}
\title{Non-equilibrium thermodynamics of stochastic systems with odd and even variables}
\author{Richard E. Spinney and
Ian J. Ford}
\affiliation{Department of Physics and Astronomy, UCL, Gower Street,
London WC1E 6BT, UK 
} 
\affiliation{London Centre for Nanotechnology, 17-19 Gordon Street
London WC1H 0AH, UK 
}

\date{December 20, 2011}
\begin{abstract}
The total entropy production of stochastic systems can be divided into three 
quantities. The first corresponds to the excess heat, whilst the second two 
comprise the house-keeping heat. We denote these two components the transient 
and generalised house-keeping heat and we obtain an integral fluctuation 
theorem for the latter, valid for all Markovian stochastic dynamics. 
A previously reported formalism is obtained when the stationary probability 
distribution is symmetric for all variables that are odd under time reversal 
which restricts consideration of directional variables such as velocity.
\end{abstract}
\maketitle

For over $100$ years the statement of the second law of thermodynamics stood 
simply as the Clausius inequality. However in recent years advances in 
technology have encouraged the thermodynamic consideration of small systems 
which has led to the generalisation of the concept of entropy production: it 
may be associated with individual dynamical realisations revealing a wealth 
of relations valid out of equilibrium. Such extensions had their origins in 
the dissipation function of Evans et al. for thermostatted systems that led 
to the Fluctuation Theorem  \cite{Evans93,Evans95,Evans02,Carberry04} with 
similar, but asymptotic relations for chaotic systems  \cite{Gallavotti95} 
which were extended to Langevin dynamics  \cite{kurchan} followed by general 
Markovian stochastic systems \cite{GCforstochastic}. Crooks and Jarzynski 
\cite{Jarzynski97,crooksoriginal,Crooks98} then derived work relations for 
a variety of dynamics which held for finite times. These were followed by 
similar generalised relations for the entropy production associated with 
transitions between stationary states  \cite{hatanosasa}, the total entropy 
production  \cite{seifertoriginal} and the heat dissipation required to 
maintain a stationary state  \cite{IFThousekeeping}. More recently the 
relationship between the latter quantities has been explored  
\cite{Jarpathintegral,Esposito07,Ge09,Ge10} resulting in a formalism 
involving a division of the total entropy change into two distinct terms, 
the adiabatic and non-adiabatic entropy productions 
\cite{adiabaticnonadiabatic0,adiabaticnonadiabatic1,adiabaticnonadiabatic2}, 
each of which obeys appropriate fluctuation relations and which map onto the 
house-keeping and excess heats, respectively, of Oono and Paniconi 
\cite{oono}. We seek to take such a formalism and generalise its scope by 
the explicit inclusion of both even (e.g. spatial) and odd (e.g. momentum)
 variables that transform differently under time reversal. In doing so we 
define a new quantity which obeys an integral fluctuation theorem for all 
time.

Specifically, we consider the dynamics of a general set of variables 
$\textbf{\em x}=(x^1,x^2,\ldots x^n)$ that behave differently under time 
reversal such that 
$\boldsymbol{\varepsilon}\textbf{\em x}=(\varepsilon^1x^1,\varepsilon^2x^2,
\ldots \varepsilon^nx^n)$ where $\varepsilon^i=\pm1$ for even and odd 
variables $x^i$ respectively. Odd variables arise in the discussion of 
directional quantities and consequently such a consideration is essential 
when discussing velocities, from the most simple lattice Boltzmann model 
to considerations of full phase space.
The entropy production of a path of duration 
$\tau$ depends on two probabilities. The first is the path probability, 
$P^{\rm F}[\vec{\textbf{\em x}}]$, defined as the probability of the forward 
trajectory, $\vec{\textbf{\em x}}=\textbf{\em x}(t)$ for $0\leq t\leq \tau$,
 with a distribution of starting configurations, 
$P^{\rm F}(\textbf{\em x}(0),0)$, that acts as an initial condition for the 
general master equation (relevant examples arise, for example, in the 
context of full phase space \cite{Brenig,kampenfluct} and in lattice 
Boltzmann models):  
\begin{equation}
  \frac{\partial P^{\rm F}(\textbf{\em x},t)}{\partial t}=
\sum_{\textbf{\em x}'}T(\textbf{\em x}|\textbf{\em x}',\lambda^{\rm F}(t))
P^{\rm F}(\textbf{\em x}',t)
\label{master}
\end{equation}
where $T(\textbf{\em x}|\textbf{\em x}',\lambda^{\rm F}(t))$ is a matrix of 
transition rates between configurations $\textbf{\em x}'$ and 
$\textbf{\em x}$, defining the normal dynamics, parameterised by the 
forward protocol $\lambda^{\rm F}$ at time $t$. We use notation 
$T(\textbf{\em x}|\textbf{\em x})=-\sum_{\textbf{\em x}'\neq 
\textbf{\em x}}T(\textbf{\em x}'|\textbf{\em x})$ which describes the mean 
escape rate. The path probability of some sequence of $N$ transitions to 
configurations $\textbf{\em x}_i$ from $\textbf{\em x}_{i-1}$ at times 
$t_i$, such that $t_0=0$ and $t_{N+1}=\tau$, can then be computed as a 
function of transition rates and exponential waiting times
\begin{align}
P^{\rm F}[\vec{\textbf{\em x}}]
&=P^{\rm F}(\textbf{\em x}_0,0)e^{\int_{t_{0}}^{t_{1}}dt'T(\textbf{\em x}_{0}|
\textbf{\em x}_{0},\lambda^{\rm F}(t'))}\nonumber\\
&\!\!\!\!\!\!\times\prod_{i=1}^{N} T(\textbf{\em x}_{i}|\textbf{\em x}_{i-1},
\lambda^{\rm F}(t_{i}))dt_i
e^{\int_{t_{i}}^{t_{i+1}}dt'T(\textbf{\em x}_{i}|\textbf{\em x}_{i},
\lambda^{\rm F}(t'))}.
\label{pathprob}
\end{align}
We compare this probability to that of another 
trajectory $\vec{\textbf{\em x}}^*$, protocol $\lambda^*$, initial condition 
$P^{*}(\textbf{\em x}^*(0),0)$ and chosen dynamics, denoted $P^{*}$, and 
write
\begin{equation}
  A[\vec{\textbf{\em x}}]=\ln\left[{P^{\rm F}[\vec{\textbf{\em x}}]}
/{P^{*}[\vec{\textbf{\em x}}^{*}]}\right].
  \label{entform}
\end{equation}
Such a quantity may obey an integral fluctuation theorem (IFT) which may be 
derived by explicit summation over all possible paths, 
$\vec{\textbf{\em x}}$, for which $P^{\rm F}[\vec{\textbf{\em x}}]\neq 0$ 
as follows
\begin{align}
  \langle \exp{\left[-A[\vec{\textbf{\em x}}]\right]}\rangle^{\rm F}
&= \sum_{\vec{\textbf{\em x}}} P^{\rm F}[\vec{\textbf{\em x}}]
\exp{\left[-A[\vec{\textbf{\em x}}]\right]}=\sum_{\vec{\textbf{\em x}}} 
P^{\rm F}[\vec{\textbf{\em x}}]\frac{P^{*}[\vec{\textbf{\em x}}^{*}]}
{P^{\rm F}[\vec{\textbf{\em x}}]}\nonumber\\
  &=\sum_{\vec{\textbf{\em x}}^{*}} P^{*}[\vec{\textbf{\em x}}^{*}]=1.
\label{IFT}
\end{align}
We assume a one to one mapping between $\vec{\textbf{\em x}}$ 
and $\vec{\textbf{\em x}}^{*}$ (a condition equivalent to a Jacobian of 
unity in the transformation) so that we may consider the summation over 
$\vec{\textbf{\em x}}^{*}$ to be equivalent to that over 
$\vec{\textbf{\em x}}$. We also require that  
$P^{*}[\vec{\textbf{\em x}}^{*}]= 0$ for all 
$P^{\rm F}[\vec{\textbf{\em x}}]= 0$ such that the final summation 
contains all possible paths $\vec{\textbf{\em x}}^{*}$, meaning the 
required normalisation of $P^{*}[\vec{\textbf{\em x}}^{*}]$ then yields 
the result of unity. A key result is the implication 
$\langle A[\vec{\textbf{\em x}}]\rangle^{\rm F}\geq 0$ by Jensen's 
inequality.
 
A common choice for $P^{*}$, and that used to construct the total entropy 
production, is that of the normal dynamics under the reversed protocol, 
denoted $P^{*}=P^{\rm R}$. Given the specification of the normal dynamics 
we point out that all further specifications, including the choice of 
protocol, can be systematically derived from the appropriate path 
transformation $\vec{\textbf{\em x}}^{*}$ which we must choose carefully 
in conjunction with the dynamics so as to obey the above conditions. At 
this point we must be clear that given a transition 
$\textbf{\em x}\to \textbf{\em x}'$ under the normal dynamics, the 
transition $\textbf{\em x}'\to\textbf{\em x}$ is not, in general, 
possible under those same dynamics. Explicitly, we can construct models 
such that $T(\textbf{\em x}'|\textbf{\em x})\neq 0$ whilst 
$T(\textbf{\em x}|\textbf{\em x}')= 0$ (as an intuitive example: 
Hamiltonian dynamics cannot produce a negative positional step whilst 
the velocity is positive). The correct path, $\vec{\textbf{\em x}}^{*}$, 
to consider is the time reversed trajectory proper which includes a 
reversal of sign for all odd variables. This is the choice 
$\textbf{\em x}^*(t)=\textbf{\em x}^{\dagger}(t)=\boldsymbol{\varepsilon}
\textbf{\em x}(\tau\!-\!t)$ and it satisfies the condition  
$P^{*}[\vec{\textbf{\em x}}^{*}]=
P^{\rm R}[\vec{\textbf{\em x}}^{\dagger}]= 0$ for all 
$P^{\rm F}[\vec{\textbf{\em x}}]= 0$ required for an IFT. The reversed 
protocol $\lambda^{*}=\lambda^{\rm R}$ may  be similarly obtained from 
the forward protocol, which may be treated as an even
 dynamical variable, meaning it transforms to yield  
$\lambda^{*}(t)=\varepsilon\lambda^{\rm F}(\tau\!-\!t)
=\lambda^{\rm F}(\tau\!-\!t)=\lambda^{\rm R}(t)$. And finally we require 
the choice of initial 
condition for the reverse path. This may be informed physically: we seek 
to characterise the irreversibility of the forward path and so initiate 
the reverse behaviour by time reversing the coordinates, 
$\textbf{\em x}(\tau)$, and distribution, $P^{\rm F}(\textbf{\em x}
(\tau),\tau)$, at the end of the forward process and evolve forward in 
time from there. The distribution can also be found by applying the 
transformation rules used to obtain the trajectory 
$\vec{\textbf{\em x}}^{\dagger}$ from $\vec{\textbf{\em x}}$  such that  
$P^{*}(\textbf{\em x}^{*}(0),0)=P^{R}(\textbf{\em x}^{\dagger}(0),0)
=\boldsymbol{\hat{\varepsilon}}P^{\rm F}(\boldsymbol{\varepsilon}
\textbf{\em x}(\tau),\tau)=P^{\rm F}(\boldsymbol{\varepsilon}
\boldsymbol{\varepsilon}
\textbf{\em x}(\tau),\tau)=P^{\rm F}(\textbf{\em x}(\tau),\tau)$ where 
$\boldsymbol{\hat{\varepsilon}}$ denotes the time reversal operation on 
the distribution. In this instance the path probability is therefore
\begin{align}
P^{\rm R}[\vec{\textbf{\em x}}^{\dagger}]
&=P^{\rm R}(\textbf{\em x}^{\dagger}_0,0)e^{\int_{t_{0}}^{t_{1}}dt'
T(\textbf{\em x}^{\dagger}_{0}|\textbf{\em x}^{\dagger}_{0},
\lambda^{\rm R}(t'))}\nonumber\\
&\!\!\!\!\!\!\!\!\!\!\!\!\times\prod_{i=1}^{N} T(\textbf{\em x}^{\dagger}_{i}|
\textbf{\em x}^{\dagger}_{i-1},\lambda^{\rm R}(t_{i}))dt_i
e^{\int_{t_{i}}^{t_{i+1}}dt'
T(\textbf{\em x}^{\dagger}_{i}|\textbf{\em x}^{\dagger}_{i},
\lambda^{\rm R}(t'))}.
\end{align}
We have 
$\textbf{\em x}_i^{\dagger}=\boldsymbol{\varepsilon}\textbf{\em x}_{N-i}$ 
so we may rearrange to give
\begin{align}
&P^{\rm R}[\vec{\textbf{\em x}}^{\dagger}]=P^{\rm F}(\textbf{\em x}_N,\tau)
e^{\int_{t_{N}}^{t_{N+1}}dt'T(\boldsymbol{\varepsilon}\textbf{\em x}_{0}|
\boldsymbol{\varepsilon}\textbf{\em x}_{0},\lambda^{\rm R}(t'))}\\
&\!\!\!\!\!\!\!\!\!\times\prod_{i=1}^{N} e^{\int_{t_{N\!-\!i}}^{t_{N\!-\!i\!+\!1}}dt'
T(\boldsymbol{\varepsilon}
\textbf{\em x}_{i}|\boldsymbol{\varepsilon}
\textbf{\em x}_{i},\lambda^{\rm R}(t'))}
T(\boldsymbol{\varepsilon}\textbf{\em x}_{i\!-\!1}|
\boldsymbol{\varepsilon}\textbf{\em x}_{i},
\lambda^{\rm R}(t_{N\!-\!i\!+\!1}))dt_i\nonumber.
\end{align}
We then perform a change of variable $t'\to \tau-t'$ and use 
$\lambda^{\rm R}(t_i)=\lambda^{\rm F}(t_{N-i+1})$ such that
\begin{align}
&P^{\rm R}[\vec{\textbf{\em x}}^{\dagger}]=P^{\rm F}(\textbf{\em x}_N,\tau)
e^{-\int_{t_{1}}^{t_{0}}dt'T(\boldsymbol{\varepsilon}\textbf{\em x}_{0}|
\boldsymbol{\varepsilon}\textbf{\em x}_{0},\lambda^{\rm F}(t'))}\\
&\times\prod_{i=1}^{N} e^{-\int_{t_{i+1}}^{t_{i}}dt'T(\boldsymbol{\varepsilon}
\textbf{\em x}_{i}|\boldsymbol{\varepsilon}\textbf{\em x}_{i},
\lambda^{\rm F}(t'))}
T(\boldsymbol{\varepsilon}\textbf{\em x}_{i-1}|
\boldsymbol{\varepsilon}\textbf{\em x}_{i},
\lambda^{\rm F}(t_{i}))dt_i\nonumber.
\end{align}
A comparison of $P^{\rm F}[\textbf{\em x}]$ and 
$P^{\rm R}[\textbf{\em x}^{\dagger}]$ characterises the irreversibility of 
the forward path and defines the total entropy production (using units 
$k_B=1$)
\begin{align}
  \Delta S_{\rm tot}&=\ln{{P^{\rm F}[\vec{\textbf{\em x}}]}}-\ln{P^{\rm R}[
\vec{\textbf{\em x}}^{\dagger}]}\nonumber\\
  &=\ln{\!\frac{P^{\rm F}(\textbf{\em x}_0,0)}
{P^{\rm F}(\textbf{\em x}_N,\tau)}}+
\sum_{i=0}^{N}\ln{\frac{e^{\int_{t_{i}}^{t_{i+1}}dt'T(\textbf{\em x}_{i}|
\textbf{\em x}_{i},
\lambda^{\rm F}(t'))}}{e^{\int_{t_{i}}^{t_{i+1}}dt'T(\boldsymbol{\varepsilon}
\textbf{\em x}_{i}|\boldsymbol{\varepsilon}\textbf{\em x}_{i},
\lambda^{\rm F}(t'))}}}\nonumber\\
&+\sum_{i=1}^{N}\ln{\!\frac{T(\textbf{\em x}_{i}|\textbf{\em x}_{i-1},
\lambda^{\rm F}(t_i))}{T(\boldsymbol{\varepsilon}\textbf{\em x}_{i-1}|
\boldsymbol{\varepsilon}\textbf{\em x}_{i},\lambda^{\rm F}(t_i))}}
\label{Stot}
\end{align}
which by its definition and Eq.~(\ref{IFT}) obeys  \cite{seifertoriginal}
\begin{equation}
  \langle \exp{[-\Delta S_{\rm tot}]}\rangle^{\rm F}=1.
\end{equation}
We find that this form of $\Delta S_{\rm tot}$ is more complicated than 
previous descriptions  \cite{Harris07,adiabaticnonadiabatic0} unless 
$\boldsymbol{\varepsilon}\textbf{\em x}=\textbf{\em x}$.  
Note that if detailed balance holds, such that $P^{\rm eq}(\textbf{\em x})
T(\textbf{\em x}'|\textbf{\em x})=P^{\rm eq}(\boldsymbol{\varepsilon}
\textbf{\em x}')T(\boldsymbol{\varepsilon}\textbf{\em x}|\boldsymbol{
\varepsilon}\textbf{\em x}')$, we expect $P^{\rm eq}$, the equilibrium 
state for a given $\lambda^{\rm F}(t)$, to satisfy $P^{\rm eq}(
\textbf{\em x})=P^{\rm eq}(\boldsymbol{\varepsilon}\textbf{\em x})$ due 
to  time-reversal invariance, along with $T(\textbf{\em x}|\textbf{\em x})
=T(\boldsymbol{\varepsilon}\textbf{\em x}|\boldsymbol{\varepsilon}
\textbf{\em x})$. For a system in equilibrium, we therefore conclude that 
$\Delta S_{\rm tot}=0$ for all paths.

Next we consider alternative specifications of $P^{*}$. We consider the 
adjoint dynamics which lead to the same stationary state, $P^{\rm st}
(\textbf{\em x},\lambda^{\rm F}(t))$, as the normal dynamics, but 
generate flux of the opposite sign in that stationary state. It can 
be shown \cite{adiabaticnonadiabatic0,Harris07,Jarpathintegral} that this 
requires an adjoint transition rate matrix $T^{\rm ad}$ described by
\begin{equation}
  T^{\rm ad}(\textbf{\em x}|\textbf{\em x}',\lambda^{\rm F}(t))
=T(\textbf{\em x}'|\textbf{\em x},
\lambda^{\rm F}(t))\frac{P^{\rm st}(\textbf{\em x},\lambda^{\rm F}(t))}{
P^{\rm st}(\textbf{\em x}',\lambda^{\rm F}(t))}.
\label{adj1}
\end{equation}
However, in the same way that the normal dynamics may 
not, in general, permit transitions $\textbf{\em x}'\to\textbf{\em x}$ or 
$\boldsymbol{\varepsilon}\textbf{\em x}\to 
\boldsymbol{\varepsilon}\textbf{\em x}'$, 
similarly the adjoint dynamics may not, in general, permit transitions 
$\textbf{\em x}\to\textbf{\em x}'$ or 
$\boldsymbol{\varepsilon}\textbf{\em x}'\to 
\boldsymbol{\varepsilon}\textbf{\em x}$. Thus we must consider the 
representation of the adjoint dynamics as either Eq.~(\ref{adj1}) or
\begin{equation}
   \!\!\!\!\!T^{\rm ad}(\boldsymbol{\varepsilon}\textbf{\em x}'|\boldsymbol{\varepsilon}
\textbf{\em x},\lambda^{\rm F}(t))=T(\boldsymbol{\varepsilon}\textbf{\em x}|
\boldsymbol{\varepsilon}\textbf{\em x}',\lambda^{\rm F}(t))\!\frac{P^{\rm st}(
\boldsymbol{\varepsilon}\textbf{\em x}',\lambda^{\rm F}(t))}{P^{\rm st}(
\boldsymbol{\varepsilon}\textbf{\em x},\lambda^{\rm F}(t))}
\label{adj2}
\end{equation}
 depending on the specific transition being considered. Explicitly, when 
choosing $P^{*}[\vec{\textbf{\em x}}^*]$, we should not consider 
$P^{\rm ad}[\vec{\textbf{\em x}}]$ or 
$P^{\rm ad}[\vec{\textbf{\em x}}^{\dagger}]$ since these might violate the 
required condition $P^{*}[\vec{\textbf{\em x}}^{*}]= 0$ for all 
$P^{\rm F}[\vec{\textbf{\em x}}]= 0$, required for an IFT.

Under the adjoint dynamics, however, an appropriate transformation of 
$\vec{\textbf{\em x}}$ is 
 $\textbf{\em x}^{*}(t)=
\textbf{\em x}^{\rm R}(t)=\textbf{\em x}(\tau\!-\!t)$. Applying the 
transformation rules used to obtain $\vec{\textbf{\em x}}^{\rm R}$ 
yields the reverse protocol as before  
$\lambda^{*}(t)=\lambda^{\rm F}(\tau\!-\!t)=\lambda^{\rm R}(t)$ and the 
initial distribution $P^{*}(\textbf{\em x}^{*}(0),0)=
P^{\rm ad,R}(\textbf{\em x}^{\rm R}(0),0)=
P^{\rm F}(\textbf{\em x}(\tau),\tau)$. The path probability is then
\begin{align}
&P^{\rm ad,R}[\vec{\textbf{\em x}}^{\rm R}]
=P^{\rm ad,R}(\textbf{\em x}^{\rm R}_0,0)
e^{\int_{t_{0}}^{t_{1}}dt'T^{\rm ad}(\textbf{\em x}^{\rm R}_{0}|
\textbf{\em x}^{\rm R}_{0},\lambda^{\rm R}(t'))}\nonumber\\
&\!\!\!\times\prod_{i=1}^{N} T^{\rm ad}(\textbf{\em x}^{\rm R}_{i}|
\textbf{\em x}^{\rm R}_{i-1},
\lambda^{\rm R}(t_{i}))dt_ie^{\int_{t_{i}}^{t_{i+1}}dt'T^{\rm ad}(
\textbf{\em x}^{\rm R}_{i}|
\textbf{\em x}^{\rm R}_{i},\lambda^{\rm R}(t'))}\nonumber\\
&=P^{\rm F}(\textbf{\em x}_N,\tau)e^{-\int_{t_{1}}^{t_{0}}dt'
T^{\rm ad}(\textbf{\em x}_{0}|
\textbf{\em x}_{0},\lambda^{\rm F}(t'))}\\
&\!\!\!\times\prod_{i=1}^{N} e^{-\int_{t_{i+1}}^{t_{i}}dt'T^{\rm ad}(
\textbf{\em x}_{i}|\textbf{\em x}_{i},\lambda^{\rm F}(t'))}
T^{\rm ad}(\textbf{\em x}_{i-1}|\textbf{\em x}_{i},
\lambda^{\rm F}(t_{i}))dt_i\nonumber.
\end{align}
We then construct a quantity of the form given in Eq.~(\ref{entform}), 
utilise Eq.~(\ref{adj1}) and the property 
$T^{\rm ad}(\textbf{\em x}|\textbf{\em x})=
T(\textbf{\em x}|\textbf{\em x})$, valid by means of balance, to obtain
\begin{align}
  \Delta S_{\rm 1}&=\ln{{P^{\rm F}[\vec{\textbf{\em x}}]}}-\ln{P^{\rm ad,R}[
\vec{\textbf{\em x}}^{\rm R}]}\nonumber\\
  &=\ln{\!\frac{P^{\rm F}(\textbf{\em x}_0,0)}
{P^{\rm F}(\textbf{\em x}_N,\tau)}}
+\sum_{i=1}^{N}\ln{\!\frac{P^{\rm st}(\textbf{\em x}_{i},
\lambda^{\rm F}(t_i))}
{P^{\rm st}(\textbf{\em x}_{i-1},\lambda^{\rm F}(t_i))}}
\end{align}
which through its definition and Eq.~(\ref{IFT}) obeys 
\begin{equation}
  \langle \exp{[-\Delta S_{1}]}\rangle^{\rm F}=1
\end{equation}
which exists in the literature as the Hatano-Sasa relation 
\cite{hatanosasa} 
or IFT for the non-adiabatic entropy production  
\cite{adiabaticnonadiabatic0,adiabaticnonadiabatic1,
adiabaticnonadiabatic2}. 
Let us now consider, once again under the adjoint dynamics, the path 
transformation choice $\textbf{\em x}^{*}(t)=\textbf{\em x}^{\rm T}(t)
=\boldsymbol{\varepsilon}\textbf{\em x}(t)$. 
Applying the transformation rules we obtain the protocol $\lambda^{*}(t)=
\varepsilon\lambda^{\rm F}(t)=\lambda^{\rm F}(t)$ and initial distribution 
$P^{*}(\textbf{\em x}^{*}(0),0)=P^{\rm ad,F}(\textbf{\em x}^{\rm T}(0),0)=
\boldsymbol{\hat{\varepsilon}}
P^{\rm F}(\boldsymbol{\varepsilon}\textbf{\em x}(0),0)
=P^{\rm F}(\textbf{\em x}(0),0)$. The path probability for this case is 
therefore
\begin{align}
&P^{\rm ad,F}[\vec{\textbf{\em x}}^{\rm T}]=
P^{\rm ad,F}(\textbf{\em x}^{\rm T}_0,0)
e^{\int_{t_{0}}^{t_{1}}dt'
T^{\rm ad}(\textbf{\em x}^{\rm T}_{0}|\textbf{\em x}^{\rm T}_{0},
\lambda^{\rm F}(t'))}\nonumber\\
&\quad\times\prod_{i=1}^{N} 
T^{\rm ad}(\textbf{\em x}^{\rm T}_{i}|\textbf{\em x}^{\rm T}_{i-1},
\lambda^{\rm F}(t_{i}))dt_ie^{\int_{t_{i}}^{t_{i+1}}dt'
T^{\rm ad}(\textbf{\em x}^{\rm T}_{i}|
\textbf{\em x}^{\rm T}_{i},\lambda^{\rm F}(t'))}\nonumber\\
&=P^{\rm F}(\textbf{\em x}_0,0)e^{\int_{t_{0}}^{t_{1}}dt'T^{\rm ad}(
\boldsymbol{\varepsilon}\textbf{\em x}_{0}|
\boldsymbol{\varepsilon}\textbf{\em x}_{0},
\lambda^{\rm F}(t'))}\\
&\quad\times\prod_{i=1}^{N} 
T^{\rm ad}(\boldsymbol{\varepsilon}\textbf{\em x}_{i}|
\boldsymbol{\varepsilon}\textbf{\em x}_{i-1},\lambda^{\rm F}(t_{i}))dt_i
e^{\int_{t_{i}}^{t_{i+1}}dt'
T^{\rm ad}(\boldsymbol{\varepsilon}\textbf{\em x}_{i}|
\boldsymbol{\varepsilon}\textbf{\em x}_{i},\lambda^{\rm F}(t'))}\nonumber.
\end{align}
By Eq.~(\ref{entform}) this then allows us to define 
\begin{align}
&\Delta S_{\rm 2}=\ln{{P^{\rm F}[\vec{\textbf{\em x}}]}}-\ln{P^{\rm ad,F}
[\vec{\textbf{\em x}}^{\rm T}]}\nonumber\\
&=\sum_{i=0}^{N}\ln{\frac{e^{\int_{t_{i}}^{t_{i+1}}dt'T(\textbf{\em x}_{i}|
\textbf{\em x}_{i},\lambda^{\rm F}(t'))}}{e^{\int_{t_{i}}^{t_{i+1}}dt'
T(\boldsymbol{\varepsilon}\textbf{\em x}_{i}|
\boldsymbol{\varepsilon}\textbf{\em x}_{i},
\lambda^{\rm F}(t'))}}}\nonumber\\
  &+\sum_{i=1}^{N}\ln{\!\frac{P^{\rm st}(\boldsymbol{\varepsilon}
\textbf{\em x}_{i-1},
\lambda^{\rm F}(t_i))}{P^{\rm st}(\boldsymbol{\varepsilon}\textbf{\em x}_i,
\lambda^{\rm F}(t_i))}\frac{T(\textbf{\em x}_{i}|
\textbf{\em x}_{i-1},\lambda^{\rm F}(t_i))}{
T(\boldsymbol{\varepsilon}\textbf{\em x}_{i-1}|
\boldsymbol{\varepsilon}\textbf{\em x}_{i},
\lambda^{\rm F}(t_i))}}
\label{S2}
\end{align}
which similarly must obey 
\begin{equation}
  \langle \exp{[-\Delta S_2]}\rangle^{\rm F}=1.
  \label{S2IFT}
\end{equation}
Unlike $\Delta S_1$, the quantity $\Delta S_2$ is new in the literature. 
We must immediately recognise that 
$\Delta S_{\rm tot}\neq \Delta S_1+\Delta S_2$ differing by a quantity
\begin{equation}
  \Delta S_3=\sum_{i=1}^{N}\ln{\!{\frac{P^{\rm st}(\textbf{\em x}_{i-1},
\lambda^{\rm F}(t_i))P^{\rm st}(\boldsymbol{\varepsilon}\textbf{\em x}_{i},
\lambda^{\rm F}(t_i))}{P^{\rm st}(\textbf{\em x}_i,\lambda^{\rm F}(t_i))
P^{\rm st}(\boldsymbol{\varepsilon}\textbf{\em x}_{i-1},
\lambda^{\rm F}(t_i))}}}
\label{S3}
\end{equation}
such that $\Delta S_{\rm tot}=\Delta S_1+\Delta S_2+\Delta S_3$. If 
$\boldsymbol{\varepsilon}\textbf{\em x}=\textbf{\em x}$ then $\Delta S_3=0$ 
and $\Delta S_2$ reduces to the adiabatic entropy production appearing in 
\cite{adiabaticnonadiabatic0,adiabaticnonadiabatic1,adiabaticnonadiabatic2}. 
More importantly we must recognise that $\Delta S_{\rm tot}\!-\!\Delta S_1\!=
\!\Delta S_2\!+\!\Delta S_3=\ln{P^{\rm ad,R}[\vec{\textbf{\em x}}^{\rm R}]}-
\ln{P^{\rm R}[\vec{\textbf{\em x}}^{\rm \dagger}]}$ or $\Delta S_{\rm tot}\!-
\!\Delta S_2\!=\!\Delta S_1\!+\!\Delta S_3=
\ln{P^{\rm ad,F}[\vec{\textbf{\em x}}^{\rm T}]}-
\ln{P^{\rm R}[\vec{\textbf{\em x}}^{\rm \dagger}]}$ cannot be written in 
the form required for Eq.~(\ref{IFT}) and so do not obey an IFT and  do not 
necessarily have any bounds on the sign of their mean. We proceed by 
following the formalism of Seifert \cite{seifertoriginal,seifertprinciples}
 and write 
\begin{equation}
  \Delta S_{\rm tot}=\ln{\frac{P^{\rm F}(\textbf{\em x}(0),0)}
{P^{\rm F}(\textbf{\em x}(\tau),\tau)}}+\frac{\Delta Q}{T_{\rm env}}=
\Delta S_{\rm sys}+\frac{\Delta Q}{T_{\rm env}},
\end{equation}
where $T_{\rm env}$ is the temperature of the environment, and that of Oono 
and Paniconi, such that total heat transfer to the environment, $\Delta Q$, 
is the sum of the excess heat and house-keeping heat $\Delta Q=
\Delta Q_{\rm ex}+\Delta Q_{\rm hk}$ \cite{oono}. The house-keeping heat is 
associated with the entropy production in stationary states and arises from 
a non-equilibrium constraint that breaks detailed balance. The sum $\Delta S_2 
+\Delta S_3$ is manifestly the entropy production in the stationary state and 
since we are considering Markov systems, both $\Delta S_2$ and $\Delta S_3$ 
are only non-zero when detailed balance is broken. Hence it is sensible to 
associate $\Delta S_2 +\Delta S_3$ with the house-keeping heat such that
\begin{equation}
  \Delta Q_{\rm hk}=(\Delta S_2\!+\!\Delta S_3)T_{\rm env}.
\end{equation}
$\Delta S_1$ is zero for all trajectories in the stationary state 
consolidating the definition of the excess heat as the heat transfer 
associated with an entropy flow that exactly cancels the change in system 
entropy in the stationary state such that
\begin{equation}
  \Delta Q_{\rm ex}=(\Delta S_1\!-\!\Delta S_{\rm sys})T_{\rm env}.
\end{equation}
However, the prevailing definition of the house-keeping heat does not make 
clear its properties when the system is not in a stationary state. A reported 
formalism suggests that it is associated with the adiabatic entropy production 
which serves as a general measure of the breakage of detailed balance  
\cite{adiabaticnonadiabatic0,adiabaticnonadiabatic1,adiabaticnonadiabatic2}.  
When considering cases where 
$\boldsymbol{\varepsilon}\textbf{\em x}=\textbf{\em x}$, this is a consistent 
approach and the mean house-keeping heat obeys strict positivity requirements 
suggesting the entropy additively increases due to non-equilibrium constraints 
and a lack of detailed balance on top of that arising from relaxation. 
However, with the inclusion of odd variables this simple picture no 
longer holds, with an ambiguity illustrated by the fact that any of 
$\Delta S_2$, $\Delta S_3$ or $\Delta S_2\!+\!\Delta S_3$ could be argued to 
be a measure of the departure from detailed balance. 
 In the light of Eq.~(\ref{S2IFT}) we propose that it is sensible to
 divide the house-keeping heat into two quantities which map onto $\Delta S_2$ 
and $\Delta S_3$. It is important to observe that, on average, the rate of 
change of $\Delta S_3$ vanishes in the stationary state by means of balance: 
the path integral over an increment in $\Delta S_3$ explicitly vanishes. 
Consequently we define the `transient house-keeping heat' 
and the `generalised house-keeping heat' 
\begin{equation}
  \Delta Q_{\rm hk,T}=\Delta S_3T_{\rm env}\quad
\quad\Delta Q_{\rm hk,G}=\Delta S_2T_{\rm env}
\end{equation}
such that $\Delta Q_{\rm hk}=\Delta Q_{\rm hk,T}\!+\!\Delta Q_{\rm hk,G}$. 
Since $\langle d\Delta S_3/d\tau\rangle ^{\rm F,st}\!=\!0$, the generalised 
house-keeping heat, when averaged, has the mean properties previously 
attributed to the house-keeping heat: it describes the heat flow required to 
maintain a non-equilibrium stationary state and is rigorously non-negative. 
Our central result therefore is
\begin{equation}
\langle \exp{[-\Delta Q_{\rm hk,G}/T_{\rm env}]}\rangle^{\rm F}=1
\end{equation} 
so $\langle\Delta Q_{\rm hk,G}\rangle^{\rm F}\geq 0$ for all times, protocols 
and initial conditions. As a corollary we also state that in general
\begin{equation}
\langle \exp{[-\Delta Q_{\rm hk}/T_{\rm env}]}\rangle^{\rm F}\neq 1
\label{noIFT}
\end{equation}
providing no bounds on $\langle\Delta Q_{\rm hk}\rangle^{\rm F}$ except 
in the stationary state when $\Delta S_1=0$ and 
$\Delta Q_{\rm hk}/T_{\rm env}=\Delta S_{\rm tot}$ or generally when 
$P^{\rm st}(\boldsymbol{\varepsilon}\textbf{\em x},
\lambda^{\rm F}(t))=P^{\rm st}(\textbf{\em x},\lambda^{\rm F}(t))$. 
As such the view that the mean rate of entropy production
is the sum of two specific non-negative contributions as in 
\cite{adiabaticnonadiabatic0,adiabaticnonadiabatic1,adiabaticnonadiabatic2}, 
is incomplete. The contribution associated with
a non-equilibrium constraint requires further unravelling, particularly
when out of stationarity.
\begin{figure}
\includegraphics[width=\columnwidth,clip]{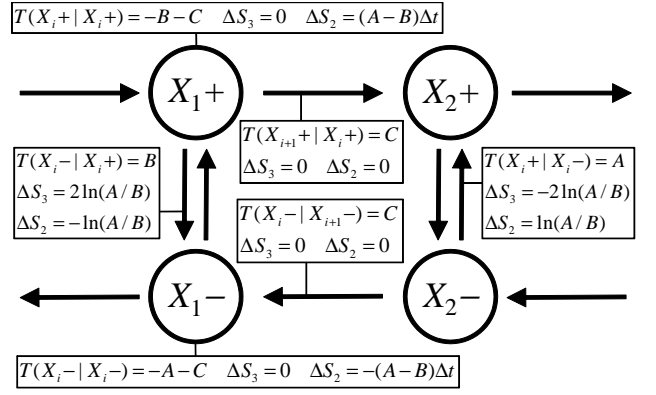}
\caption{\label{lattice}Allowed moves between positions $X_i$ and $\pm$ 
velocity states are shown by arrows, with associated rates $T$. 
Periodic boundaries allow jumps from $X_L+$ to $X_1+$ and $X_{1}-$ to 
$X_{L}-$. A given path contributes to the transient and generalised 
house-keeping heats, $T_{\rm env}\Delta S_3$ and $T_{\rm env}
\Delta S_2$, respectively, due to transitions between, and residence 
times $\Delta t$ at, each phase space point, as indicated. These 
correspond to individual terms in the summations in Eqs.~(\ref{S2}) 
and (\ref{S3}). }
\end{figure}

To explore the nature of the house-keeping heat we consider its behaviour 
in the approach to the stationary state of a simple model of particle 
dynamics on a ring. The phase space consists of $L$ identical spatial 
positions $X_{1},X_{2}\ldots X_{L}$ and two velocities labelled $+$ and 
$-$ as shown in Fig.~\ref{lattice} with the time reversal properties 
$\boldsymbol{\varepsilon}X_i\pm=X_i\mp$ necessitated by the one-way nature 
of many of the transitions. The stationary state probabilities that arise 
from these dynamics are $P^{\rm st}(X_i+)=A/(L(A+B))$ and 
$P^{\rm st}(X_i-)=B/(L(A+B))$. Any difference between the velocity 
reversal rates $A$ and $B$ gives rise to a non-equilibrium stationary 
state by providing a stationary particle current, which for $A>B$ runs 
from left to right. Such dynamics amount to a very simple lattice 
Boltzmann model. Contributions $\Delta S_2$ and $\Delta S_3$ associated 
with particle behaviour consisting of instantaneous transitions and 
waiting periods are indicated. 
We consider particle behaviour over a small time interval $dt$, and 
compute the mean entropy production rates to leading order in $dt$. 
Examining the path probability in Eq.~(\ref{pathprob}) we need only 
consider $N=0$ or $N=1$ transitions. Identifying leading order terms in 
the products of $P$, $T$, exponentiated waiting times and $\Delta S_3$ 
that make up the average of the form given in Eq.~(\ref{IFT}) yields
\begin{equation}
  \frac{d\langle\Delta S_3\rangle^{\rm F}}{dt}=\sum_{i=1}^{L}
2P(X_i+)B\ln{\frac{A}{B}}+2P(X_i-)A\ln{\frac{B}{A}}.
\label{S3dt}
\end{equation}
For non-stationary $P$ its sign is unbounded: for example if all the
 probability were uniformly distributed initially amongst the $+$ 
velocity states it would equal $2B\ln(A/B)$, whilst if it were 
distributed over the $-$ states it would be $-2A\ln(A/B)$ instead. 
Such non-zero contributions to $\Delta S_3$ require an asymmetric 
stationary state in odd variables which thus explains their absence 
when the stationary velocity distribution is assumed to be symmetric, 
such as in  overdamped Langevin descriptions (see 
\cite{IFThousekeeping} and examples in \cite{adiabaticnonadiabatic2}). 
However, in the stationary state with $P=P^{\rm st}$, $d\langle\Delta 
S_3\rangle^{\rm F}/dt$ is demonstrably equal to zero as claimed. By 
similar means 
\begin{align}
\frac{d\langle\Delta S_2\rangle^{\rm F}}{dt}=\sum_{i=1}^{L}&P(X_i+)
\left[A-B-B\ln{\frac{A}{B}}\right]\nonumber\\
&+P(X_i-)\left[B-A-A\ln{\frac{B}{A}}\right]
\label{S2dt}
\end{align}
which is positive for all positive $A$ and $B$ and reduces to 
$d\langle\Delta S_2\rangle^{\rm F,st}/dt=(A-B)^2/(A+B)$ in the 
stationary state. We note that the sum of Eqs.~(\ref{S3dt}) and 
(\ref{S2dt}) has no bound on its sign and relates to the inequality 
in Eq.~(\ref{noIFT}). Further, $d\langle\exp{[-\Delta S_{2}]}
\rangle^{\rm F}/dt=0$ and $\langle \exp{[-\Delta S_2(t=0)]}
\rangle^{\rm F}=1$ which explicitly demonstrates the expected IFT 
for any normalised $P(X_i\pm)$. Finally, we note that for $A=B$, 
all contributions vanish in detail as this corresponds to equilibrium 
where there is no entropy production.

We have extended the formalism found in 
\cite{hatanosasa,IFThousekeeping,adiabaticnonadiabatic0,adiabaticnonadiabatic1,
adiabaticnonadiabatic2} and split the total entropy production into two 
rigorously positive contributions and a third contribution which has no bounds 
on its sign. We have argued that this final quantity is, in the mean, a 
transient contribution to the house-keeping heat and it is the mean 
generalised house-keeping heat that is rigorously positive for all times. It 
is not straightforward to consolidate this with the two causes of time 
reversal asymmetry namely relaxation to the stationary state and imposed 
non-equilibrium constraints: $\Delta S_3$ exists only in the presence the 
latter, but is, in the mean, its own measure of relaxation to the stationary 
state. It could be argued that the non-adiabatic entropy production and 
Hatano-Sasa relation do not fully capture the entropy production due to 
transitions between stationary states, but associating $\Delta S_3$ with one 
or other form of entropy production is not entirely satisfactory as it occurs 
when the line between them is blurred. Nevertheless, either interpretation 
elucidates a new layer of complexity in the theory of entropy production in
 stochastic systems. Further exploration in the context of continuous 
stochastic processes is to be reported elsewhere \cite{SpinneyFord}. The 
authors acknowledge financial support from EPSRC.
\providecommand{\noopsort}[1]{}\providecommand{\singleletter}[1]{#1}%

\begin{thebibliography}{26}%
\makeatletter
\providecommand \@ifxundefined [1]{%
 \@ifx{#1\undefined}
}%
\providecommand \@ifnum [1]{%
 \ifnum #1\expandafter \@firstoftwo
 \else \expandafter \@secondoftwo
 \fi
}%
\providecommand \@ifx [1]{%
 \ifx #1\expandafter \@firstoftwo
 \else \expandafter \@secondoftwo
 \fi
}%
\providecommand \natexlab [1]{#1}%
\providecommand \enquote  [1]{``#1''}%
\providecommand \bibnamefont  [1]{#1}%
\providecommand \bibfnamefont [1]{#1}%
\providecommand \citenamefont [1]{#1}%
\providecommand \href@noop [0]{\@secondoftwo}%
\providecommand \href [0]{\begingroup \@sanitize@url \@href}%
\providecommand \@href[1]{\@@startlink{#1}\@@href}%
\providecommand \@@href[1]{\endgroup#1\@@endlink}%
\providecommand \@sanitize@url [0]{\catcode `\\12\catcode `\$12\catcode
  `\&12\catcode `\#12\catcode `\^12\catcode `\_12\catcode `\%12\relax}%
\providecommand \@@startlink[1]{}%
\providecommand \@@endlink[0]{}%
\providecommand \url  [0]{\begingroup\@sanitize@url \@url }%
\providecommand \@url [1]{\endgroup\@href {#1}{\urlprefix }}%
\providecommand \urlprefix  [0]{URL }%
\providecommand \Eprint [0]{\href }%
\providecommand \doibase [0]{http://dx.doi.org/}%
\providecommand \selectlanguage [0]{\@gobble}%
\providecommand \bibinfo  [0]{\@secondoftwo}%
\providecommand \bibfield  [0]{\@secondoftwo}%
\providecommand \translation [1]{[#1]}%
\providecommand \BibitemOpen [0]{}%
\providecommand \bibitemStop [0]{}%
\providecommand \bibitemNoStop [0]{.\EOS\space}%
\providecommand \EOS [0]{\spacefactor3000\relax}%
\providecommand \BibitemShut  [1]{\csname bibitem#1\endcsname}%
\let\auto@bib@innerbib\@empty
\bibitem [{\citenamefont {Evans}\ \emph {et~al.}(1993)\citenamefont {Evans},
  \citenamefont {Cohen},\ and\ \citenamefont {Morriss}}]{Evans93}%
  \BibitemOpen
  \bibfield  {author} {\bibinfo {author} {\bibfnamefont {D.~J.}\ \bibnamefont
  {Evans}}, \bibinfo {author} {\bibfnamefont {E.~G.~D.}\ \bibnamefont {Cohen}},
  \ and\ \bibinfo {author} {\bibfnamefont {G.~P.}\ \bibnamefont {Morriss}},\
  }\href@noop {} {\bibfield  {journal} {\bibinfo  {journal} {Phys. Rev. Lett.}\
  }\textbf {\bibinfo {volume} {71}},\ \bibinfo {pages} {2401} (\bibinfo {year}
  {1993})}\BibitemShut {NoStop}%
\bibitem [{\citenamefont {Evans}\ and\ \citenamefont
  {Searles}(1995)}]{Evans95}%
  \BibitemOpen
  \bibfield  {author} {\bibinfo {author} {\bibfnamefont {D.~J.}\ \bibnamefont
  {Evans}}\ and\ \bibinfo {author} {\bibfnamefont {D.~J.}\ \bibnamefont
  {Searles}},\ }\href@noop {} {\bibfield  {journal} {\bibinfo  {journal} {Phys.
  Rev. E}\ }\textbf {\bibinfo {volume} {52}},\ \bibinfo {pages} {5839}
  (\bibinfo {year} {1995})}\BibitemShut {NoStop}%
\bibitem [{\citenamefont {Evans}\ and\ \citenamefont
  {Searles}(2002)}]{Evans02}%
  \BibitemOpen
  \bibfield  {author} {\bibinfo {author} {\bibfnamefont {D.~J.}\ \bibnamefont
  {Evans}}\ and\ \bibinfo {author} {\bibfnamefont {D.~J.}\ \bibnamefont
  {Searles}},\ }\href@noop {} {\bibfield  {journal} {\bibinfo  {journal} {Adv.
  Phys.}\ }\textbf {\bibinfo {volume} {51}},\ \bibinfo {pages} {1529} (\bibinfo
  {year} {2002})}\BibitemShut {NoStop}%
\bibitem [{\citenamefont {Carberry}\ \emph {et~al.}(2004)\citenamefont
  {Carberry}, \citenamefont {Reid}, \citenamefont {G.M.Wang}, \citenamefont
  {Sevick}, \citenamefont {Searles},\ and\ \citenamefont {Evans}}]{Carberry04}%
  \BibitemOpen
  \bibfield  {author} {\bibinfo {author} {\bibfnamefont {D.}~\bibnamefont
  {Carberry}}, \bibinfo {author} {\bibfnamefont {J.~C.}\ \bibnamefont {Reid}},
  \bibinfo {author} {\bibnamefont {G.M.Wang}}, \bibinfo {author} {\bibfnamefont
  {E.}~\bibnamefont {Sevick}}, \bibinfo {author} {\bibfnamefont {D.~J.}\
  \bibnamefont {Searles}}, \ and\ \bibinfo {author} {\bibfnamefont {D.~J.}\
  \bibnamefont {Evans}},\ }\href@noop {} {\bibfield  {journal} {\bibinfo
  {journal} {Phys. Rev. Lett.}\ }\textbf {\bibinfo {volume} {92}},\ \bibinfo
  {pages} {140601} (\bibinfo {year} {2004})}\BibitemShut {NoStop}%
\bibitem [{\citenamefont {Gallavotti}\ and\ \citenamefont
  {Cohen}(1995)}]{Gallavotti95}%
  \BibitemOpen
  \bibfield  {author} {\bibinfo {author} {\bibfnamefont {G.}~\bibnamefont
  {Gallavotti}}\ and\ \bibinfo {author} {\bibfnamefont {E.~G.~D.}\ \bibnamefont
  {Cohen}},\ }\href@noop {} {\bibfield  {journal} {\bibinfo  {journal} {Phys.
  Rev. Lett.}\ }\textbf {\bibinfo {volume} {74}},\ \bibinfo {pages} {2694}
  (\bibinfo {year} {1995})}\BibitemShut {NoStop}%
\bibitem [{\citenamefont {Kurchan}(1998)}]{kurchan}%
  \BibitemOpen
  \bibfield  {author} {\bibinfo {author} {\bibfnamefont {J.}~\bibnamefont
  {Kurchan}},\ }\href@noop {} {\bibfield  {journal} {\bibinfo  {journal} {J.
  Phys. A: Math. Gen.}\ }\textbf {\bibinfo {volume} {31}},\ \bibinfo {pages}
  {3719} (\bibinfo {year} {1998})}\BibitemShut {NoStop}%
\bibitem [{\citenamefont {Lebowitz}\ and\ \citenamefont
  {Spohn}(1999)}]{GCforstochastic}%
  \BibitemOpen
  \bibfield  {author} {\bibinfo {author} {\bibfnamefont {J.~L.}\ \bibnamefont
  {Lebowitz}}\ and\ \bibinfo {author} {\bibfnamefont {H.}~\bibnamefont
  {Spohn}},\ }\href@noop {} {\bibfield  {journal} {\bibinfo  {journal} {J.
  Stat. Phys.}\ }\textbf {\bibinfo {volume} {95}},\ \bibinfo {pages} {333}
  (\bibinfo {year} {1999})}\BibitemShut {NoStop}%
\bibitem [{\citenamefont {Jarzynski}(1997)}]{Jarzynski97}%
  \BibitemOpen
  \bibfield  {author} {\bibinfo {author} {\bibfnamefont {C.}~\bibnamefont
  {Jarzynski}},\ }\href@noop {} {\bibfield  {journal} {\bibinfo  {journal}
  {Phys. Rev. Lett.}\ }\textbf {\bibinfo {volume} {78}},\ \bibinfo {pages}
  {2690} (\bibinfo {year} {1997})}\BibitemShut {NoStop}%
\bibitem [{\citenamefont {Crooks}(1999)}]{crooksoriginal}%
  \BibitemOpen
  \bibfield  {author} {\bibinfo {author} {\bibfnamefont {G.~E.}\ \bibnamefont
  {Crooks}},\ }\href@noop {} {\bibfield  {journal} {\bibinfo  {journal} {Phys.
  Rev. E}\ }\textbf {\bibinfo {volume} {60}},\ \bibinfo {pages} {2721}
  (\bibinfo {year} {1999})}\BibitemShut {NoStop}%
\bibitem [{\citenamefont {Crooks}(1998)}]{Crooks98}%
  \BibitemOpen
  \bibfield  {author} {\bibinfo {author} {\bibfnamefont {G.~E.}\ \bibnamefont
  {Crooks}},\ }\href@noop {} {\bibfield  {journal} {\bibinfo  {journal} {J.
  Stat. Phys.}\ }\textbf {\bibinfo {volume} {90}},\ \bibinfo {pages} {1481}
  (\bibinfo {year} {1998})}\BibitemShut {NoStop}%
\bibitem [{\citenamefont {Hatano}\ and\ \citenamefont
  {Sasa}(2001)}]{hatanosasa}%
  \BibitemOpen
  \bibfield  {author} {\bibinfo {author} {\bibfnamefont {T.}~\bibnamefont
  {Hatano}}\ and\ \bibinfo {author} {\bibfnamefont {S.}~\bibnamefont {Sasa}},\
  }\href@noop {} {\bibfield  {journal} {\bibinfo  {journal} {Phys. Rev. Lett.}\
  }\textbf {\bibinfo {volume} {86}},\ \bibinfo {pages} {3463} (\bibinfo {year}
  {2001})}\BibitemShut {NoStop}%
\bibitem [{\citenamefont {Seifert}(2005)}]{seifertoriginal}%
  \BibitemOpen
  \bibfield  {author} {\bibinfo {author} {\bibfnamefont {U.}~\bibnamefont
  {Seifert}},\ }\href@noop {} {\bibfield  {journal} {\bibinfo  {journal} {Phys.
  Rev. Lett.}\ }\textbf {\bibinfo {volume} {95}},\ \bibinfo {pages} {040602}
  (\bibinfo {year} {2005})}\BibitemShut {NoStop}%
\bibitem [{\citenamefont {Speck}\ and\ \citenamefont
  {Seifert}(2005)}]{IFThousekeeping}%
  \BibitemOpen
  \bibfield  {author} {\bibinfo {author} {\bibfnamefont {T.}~\bibnamefont
  {Speck}}\ and\ \bibinfo {author} {\bibfnamefont {U.}~\bibnamefont
  {Seifert}},\ }\href@noop {} {\bibfield  {journal} {\bibinfo  {journal} {J.
  Phys. A: Math. Gen.}\ }\textbf {\bibinfo {volume} {38}},\ \bibinfo {pages}
  {L581} (\bibinfo {year} {2005})}\BibitemShut {NoStop}%
\bibitem [{\citenamefont {Chernyak}\ \emph {et~al.}(2006)\citenamefont
  {Chernyak}, \citenamefont {Chertkov},\ and\ \citenamefont
  {Jarzynski}}]{Jarpathintegral}%
  \BibitemOpen
  \bibfield  {author} {\bibinfo {author} {\bibfnamefont {V.~Y.}\ \bibnamefont
  {Chernyak}}, \bibinfo {author} {\bibfnamefont {M.}~\bibnamefont {Chertkov}},
  \ and\ \bibinfo {author} {\bibfnamefont {C.}~\bibnamefont {Jarzynski}},\
  }\href@noop {} {\bibfield  {journal} {\bibinfo  {journal} {J. Stat. Mech.
  \textup{P08001}}\ } (\bibinfo {year} {2006})}\BibitemShut {NoStop}%
\bibitem [{\citenamefont {Esposito}\ \emph {et~al.}(2007)\citenamefont
  {Esposito}, \citenamefont {Harbola},\ and\ \citenamefont
  {Mukamel}}]{Esposito07}%
  \BibitemOpen
  \bibfield  {author} {\bibinfo {author} {\bibfnamefont {M.}~\bibnamefont
  {Esposito}}, \bibinfo {author} {\bibfnamefont {U.}~\bibnamefont {Harbola}}, \
  and\ \bibinfo {author} {\bibfnamefont {S.}~\bibnamefont {Mukamel}},\ }\href
  {\doibase 10.1103/PhysRevE.76.031132} {\bibfield  {journal} {\bibinfo
  {journal} {Phys. Rev. E}\ }\textbf {\bibinfo {volume} {76}},\ \bibinfo
  {pages} {031132} (\bibinfo {year} {2007})}\BibitemShut {NoStop}%
\bibitem [{\citenamefont {Ge}(2009)}]{Ge09}%
  \BibitemOpen
  \bibfield  {author} {\bibinfo {author} {\bibfnamefont {H.}~\bibnamefont
  {Ge}},\ }\href {\doibase 10.1103/PhysRevE.80.021137} {\bibfield  {journal}
  {\bibinfo  {journal} {Phys. Rev. E}\ }\textbf {\bibinfo {volume} {80}},\
  \bibinfo {pages} {021137} (\bibinfo {year} {2009})}\BibitemShut {NoStop}%
\bibitem [{\citenamefont {Ge}\ and\ \citenamefont {Qian}(2010)}]{Ge10}%
  \BibitemOpen
  \bibfield  {author} {\bibinfo {author} {\bibfnamefont {H.}~\bibnamefont
  {Ge}}\ and\ \bibinfo {author} {\bibfnamefont {H.}~\bibnamefont {Qian}},\
  }\href {\doibase 10.1103/PhysRevE.81.051133} {\bibfield  {journal} {\bibinfo
  {journal} {Phys. Rev. E}\ }\textbf {\bibinfo {volume} {81}},\ \bibinfo
  {pages} {051133} (\bibinfo {year} {2010})}\BibitemShut {NoStop}%
\bibitem [{\citenamefont {Esposito}\ and\ \citenamefont {Van~den
  Broeck}(2010{\natexlab{a}})}]{adiabaticnonadiabatic0}%
  \BibitemOpen
  \bibfield  {author} {\bibinfo {author} {\bibfnamefont {M.}~\bibnamefont
  {Esposito}}\ and\ \bibinfo {author} {\bibfnamefont {C.}~\bibnamefont {Van~den
  Broeck}},\ }\href@noop {} {\bibfield  {journal} {\bibinfo  {journal} {Phys.
  Rev. Lett.}\ }\textbf {\bibinfo {volume} {104}},\ \bibinfo {pages} {090601}
  (\bibinfo {year} {2010}{\natexlab{a}})}\BibitemShut {NoStop}%
\bibitem [{\citenamefont {Esposito}\ and\ \citenamefont {Van~den
  Broeck}(2010{\natexlab{b}})}]{adiabaticnonadiabatic1}%
  \BibitemOpen
  \bibfield  {author} {\bibinfo {author} {\bibfnamefont {M.}~\bibnamefont
  {Esposito}}\ and\ \bibinfo {author} {\bibfnamefont {C.}~\bibnamefont {Van~den
  Broeck}},\ }\href@noop {} {\bibfield  {journal} {\bibinfo  {journal} {Phys.
  Rev. E}\ }\textbf {\bibinfo {volume} {82}},\ \bibinfo {pages} {011143}
  (\bibinfo {year} {2010}{\natexlab{b}})}\BibitemShut {NoStop}%
\bibitem [{\citenamefont {Van~den Broeck}\ and\ \citenamefont
  {Esposito}(2010)}]{adiabaticnonadiabatic2}%
  \BibitemOpen
  \bibfield  {author} {\bibinfo {author} {\bibfnamefont {C.}~\bibnamefont
  {Van~den Broeck}}\ and\ \bibinfo {author} {\bibfnamefont {M.}~\bibnamefont
  {Esposito}},\ }\href@noop {} {\bibfield  {journal} {\bibinfo  {journal}
  {Phys. Rev. E}\ }\textbf {\bibinfo {volume} {82}},\ \bibinfo {pages} {011144}
  (\bibinfo {year} {2010})}\BibitemShut {NoStop}%
\bibitem [{\citenamefont {Oono}\ and\ \citenamefont {Paniconi}(1998)}]{oono}%
  \BibitemOpen
  \bibfield  {author} {\bibinfo {author} {\bibfnamefont {Y.}~\bibnamefont
  {Oono}}\ and\ \bibinfo {author} {\bibfnamefont {M.}~\bibnamefont
  {Paniconi}},\ }\href@noop {} {\bibfield  {journal} {\bibinfo  {journal}
  {Prog. Theor. Phys. Suppl.}\ }\textbf {\bibinfo {volume} {130}},\ \bibinfo
  {pages} {29} (\bibinfo {year} {1998})}\BibitemShut {NoStop}%
\bibitem [{\citenamefont {Brenig}\ and\ \citenamefont {Van~den
  Broeck}(1980)}]{Brenig}%
  \BibitemOpen
  \bibfield  {author} {\bibinfo {author} {\bibfnamefont {L.}~\bibnamefont
  {Brenig}}\ and\ \bibinfo {author} {\bibfnamefont {C.}~\bibnamefont {Van~den
  Broeck}},\ }\href@noop {} {\bibfield  {journal} {\bibinfo  {journal} {Phys.
  Rev. A}\ }\textbf {\bibinfo {volume} {21}},\ \bibinfo {pages} {1039}
  (\bibinfo {year} {1980})}\BibitemShut {NoStop}%
\bibitem [{\citenamefont {van Kampen}(1976)}]{kampenfluct}%
  \BibitemOpen
  \bibfield  {author} {\bibinfo {author} {\bibfnamefont {N.~G.}\ \bibnamefont
  {van Kampen}},\ }\href@noop {} {\bibfield  {journal} {\bibinfo  {journal}
  {AIP Conference Proceedings}\ }\textbf {\bibinfo {volume} {27}},\ \bibinfo
  {pages} {153} (\bibinfo {year} {1976})}\BibitemShut {NoStop}%
\bibitem [{\citenamefont {Harris}\ and\ \citenamefont
  {Sch{\"u}tz}(2007)}]{Harris07}%
  \BibitemOpen
  \bibfield  {author} {\bibinfo {author} {\bibfnamefont {R.~J.}\ \bibnamefont
  {Harris}}\ and\ \bibinfo {author} {\bibfnamefont {G.~M.}\ \bibnamefont
  {Sch{\"u}tz}},\ }\href@noop {} {\bibfield  {journal} {\bibinfo  {journal} {J.
  Stat. Mech. \textup{P07020}}\ } (\bibinfo {year} {2007})}\BibitemShut
  {NoStop}%
\bibitem [{\citenamefont {{Seifert, U.}}(2008)}]{seifertprinciples}%
  \BibitemOpen
  \bibfield  {author} {\bibinfo {author} {\bibnamefont {{Seifert, U.}}},\
  }\href@noop {} {\bibfield  {journal} {\bibinfo  {journal} {Eur. Phys. J. B}\
  }\textbf {\bibinfo {volume} {64}},\ \bibinfo {pages} {423} (\bibinfo {year}
  {2008})}\BibitemShut {NoStop}%
\bibitem [{\citenamefont {Spinney}\ and\ \citenamefont {Ford}()}]{SpinneyFord}%
  \BibitemOpen
  \bibfield  {author} {\bibinfo {author} {\bibfnamefont {R.~E.}\ \bibnamefont
  {Spinney}}\ and\ \bibinfo {author} {\bibfnamefont {I.~J.}\ \bibnamefont
  {Ford}},\ }\href@noop {} {\bibinfo  {journal} {In preparation}\ }\BibitemShut
  {NoStop}%
\end{thebibliography}
\end{document}